\documentclass[aps,pre,twocolumn,groupedaddress,showpacs]{revtex4-1}

\usepackage{natbib}
\usepackage{graphics}
\usepackage{amsfonts}
\usepackage{amssymb}
\usepackage{amsmath}

\begin{document}

\title{Further details on the phase diagram of hard ellipsoids of revolution}

\author{Gustavo Bautista-Carbajal}
\affiliation{Departamento de F\'isica, Universidad Aut\'onoma Metropolitana-Iztapalapa, 09340, M\'exico, Distrito Federal,
M\'{e}xico,}
\affiliation{Academia de Matem\'aticas, Universidad Aut\'onoma de la Ciudad de M\'exico, 07160, M\'exico, Distrito Federal,
M\'exico. }

\author{Arturo Moncho-Jord\'a}
\affiliation{Departamento de F\'isica Aplicada, Facultad de Ciencias, Universidad de Granada, Campus de Fuentenueva,  18071 Granada, Spain}

\author{Gerardo Odriozola}
\email{godriozo@imp.mx} 
\affiliation{Programa de Ingenier\'{\i}a
Molecular, Instituto Mexicano del Petr\'{o}leo, Eje Central
L\'{a}zaro C\'ardenas 152, 07730, M\'{e}xico, Distrito Federal,
M\'{e}xico.}

\date{\today}

\begin{abstract}
In recent work we revisited the phase diagram of hard ellipsoids of revolution (spheroids) by means of replica exchange Monte Carlo simulations. This was done by setting random initial configurations, and allows to confirm the formation of sm2 crystal structures at high densities [Phys. Rev. E 75, 020402 (2007)] for large anisotropies and stretched-fcc for small anisotropies. In this work we employed the same technique but setting the starting cells as sm2 crystal structures having the maximum known packing density [Phys. Rev. Lett. 92, 255506 (2004)]. This procedure yields a very rich behavior for quasi-spherical oblates and prolates. These systems, from low to high pressures, show the following phases: isotropic fluid, plastic solid, stretched-fcc solid, and sm2 solid. The first three transitions are first order, whereas the last one is a subtle, probably high order transition. This picture is consistent with the fact of having the sm2 structure capable of producing the maximally achievable density.
\end{abstract}


\maketitle

\section{Introduction}

During quite a long time it was assumed that the maximum achievable density of hard ellipsoids was given by the stretched face cubic centered (sfcc), with a volume fraction of $\varphi_m = \pi/\sqrt{18} \approx 0.7405$. However, after the observation of certain crystal structures capable of surpassing this threshold~\cite{Donev04a}, the high density region of the original phase diagram~\cite{Frenkel84,Frenkel85} has been significantly modified~\cite{Pfleiderer07,Radu09,Odriozola12}. Naturally, the updated phase diagram must include the crystal structures which, under large compressions, produce the maximally achievable densities. In particular, Pfleiderer and Schilling~\cite{Pfleiderer07} found that a family of crystals named sm2 (simple monoclinic with two orientations) showed smaller free energies than that of the sfcc for ellipsoids having large asymmetries, whereas the opposite was found for ellipsoids with small asymmetries~\cite{Radu09}. On the other hand, the structures given by Donev et.~al.~showing 
the maximally known achievable densities are particular cases of the sm2 family~\cite{Pfleiderer07,Radu09}.

At present, the phase diagram of hard ellipsoids with large asymmetry shows consistency between the free energy predictions and the structure showing the maximally achievable density. From low to high densities, there is an isotropic fluid, a nematic fluid, and finally a sm2 crystal structure which would yield the densest structure under infinite pressure. Conversely, the low asymmetry region still looks incomplete. That is, there is an isotropic fluid, a plastic crystal~\cite{Frenkel85}, and a sfcc structure of parallel ellipsoids which cannot be compressed to reach the maximum packing fraction (parallel ellipsoids cannot exceed the fcc density limit~\cite{TorquatoRev}). Consequently, the sfcc structure of parallel ellipsoids must somehow distort to increase its density under extreme compression, or suffer another phase transition to produce the sm2 structure. 

In recent work we revisited the phase diagram of hard ellipsoids of revolution by means of replica exchange Monte Carlo simulations~\cite{Odriozola12}. This was done by setting random initial conditions, and the results confirmed the spontaneous formation of sm2 crystal structures at high densities and relatively large anisotropies, and parallel sfcc structures for small anisotropies. In the present work we employ the same technique but setting the starting cells as perfect sm2 crystal structures having the maximum known packing density~\cite{Donev04a}. This is done with the hope that these structures are indeed the ones that reach the maximum packing density. If so, the obtained results should correspond to equilibrium. Anyway, a direct comparison with the pressure-density curves from random initial conditions is possible. If curves match, equilibrium would be the case. On the contrary, the method will be certainly failing in the sense that the ergodic hypothesis is violated for the given conditions. 

The setting of perfect sm2 structures as initial conditions allows us to capture a very high pressure sfcc-sm2 transition for small anisotropies. In this region, we find that the sm2 structure holds only at very high pressures, turning first into a parallel sfcc structure and then into a plastic-solid when decreasing pressure and density. According to our results, obtained for small systems, the sfcc-sm2 transition is not first order. The existence of this sfcc-sm2 transition occurs only above the plastic solid phase, for both, oblates and prolates, whereas the sfcc structure region vanishes at large anisotropies. 

The paper is organized in four sections. Following this brief introduction, the second section describes the employed models and methods. In a third section, the high pressure phase transition for the small anisotropic ellipsoids is shown, together with a comparison between the results obtained by starting from sm2 structures and from random initial configurations. Some remarks and conclusions are given in a final section.    

\section{Models and methods}

\subsection{Hard ellipsoids model}

In order to perform a simulation of hard ellipsoids, we need an efficient way to avoid particle overlapping. For this purpose, we use an analytical approach for the exact hard ellipsoids contact distance. The expression is based on the Berne and Pechukas~\cite{Berne72} closest approach distance (also called hard Gaussian overlap~\cite{Varga02}), and includes a corrective term to improve precision. This term was introduced by Rickayzen to fix the known T-shape Berne and Pechukas mismatch. The Rickayzen-Berne-Pechukas (RBP) expression reads~\cite{Rickayzen98}
\begin{equation}\label{RBP1}
\sigma_{RBP} = \frac{\sigma_{\bot}}{\sqrt{ 1 - \frac{1}{2} \chi \big [ A^{+} + A^{-} \big ] + \big ( 1 - \chi ) \chi' \big [ A^{+} A^{-} \big ]^{\gamma} }},
\end{equation}
being
\begin{equation}\label{RBP2}
A^{\pm} = \frac{ ( \hat{\mathbf{r}} \cdot \hat{\mathbf{u}}_{i} \pm \hat{\mathbf{r}} \cdot \hat{\mathbf{u}}_{j} )^{2} }{ 1 \pm \chi \hat{\mathbf{u}}_{i} \cdot \hat{\mathbf{u}}_{j} },
\end{equation}
\begin{equation}\label{RP2}
\chi = \frac{ \sigma_{\|}^{2} - \sigma_{\bot}^{2} }{ \sigma_{\|}^{2} + \sigma_{\bot}^{2} } \ \ , \ \ \chi' = \bigg ( \frac{ \sigma_{\|} - \sigma_{\bot} }{ \sigma_{\|} + \sigma_{\bot} } \bigg )^{2}.
\end{equation}
Here, $\sigma_{\|}$ and $\sigma_{\bot}$ are the parallel and perpendicular diameters with respect to the ellipsoid axis of revolution, respectively. We define the aspect ratio of the ellipsoids as $\alpha=\sigma_{\|}/\sigma_{\bot}$, such that $\alpha > 1$ corresponds to prolates, and $\alpha < 1$ to oblates. Analogously, the maximum aspect ratio is defined as $\delta = \sigma_{max}/\sigma_{min}$, where $\sigma_{max}$ and $\sigma_{min}$ are the maximum and minimum axes, respectively. $\hat{\mathbf{u}}_{i}$ and $\hat{\mathbf{u}}_{j}$ are unit vectors along the axis of revolution of each particle, and $\hat{\mathbf{r}}$ is the unit vector along the line joining the geometric particle centers. Finally, $\gamma$ is introduced~\cite{GuevaraHE} to further approach to the exact Perram and Wertheim numerical solution~\cite{Perram84,Perram85}. $\gamma$ values are given in reference~\cite{GuevaraHE}. The average difference between the analytical approach and the exact numerical solution is always below $0.8\%$ for $0.2 
\leq \alpha \leq 5$, for a collection of $10^8$ random configurations (varying $\hat{\mathbf{r}}$, $\hat{\mathbf{u}}_{i}$, and $\hat{\mathbf{u}}_{j}$). The equations of state corresponding to the RBP and PW were also compared showing no practical differences for oblates with $\delta = 5$~\cite{GuevaraHE}.  

\subsection{Donev-sm2 structure}
\label{section-Donev}

\begin{figure} 
\center\resizebox{0.48\textwidth}{!}{\includegraphics{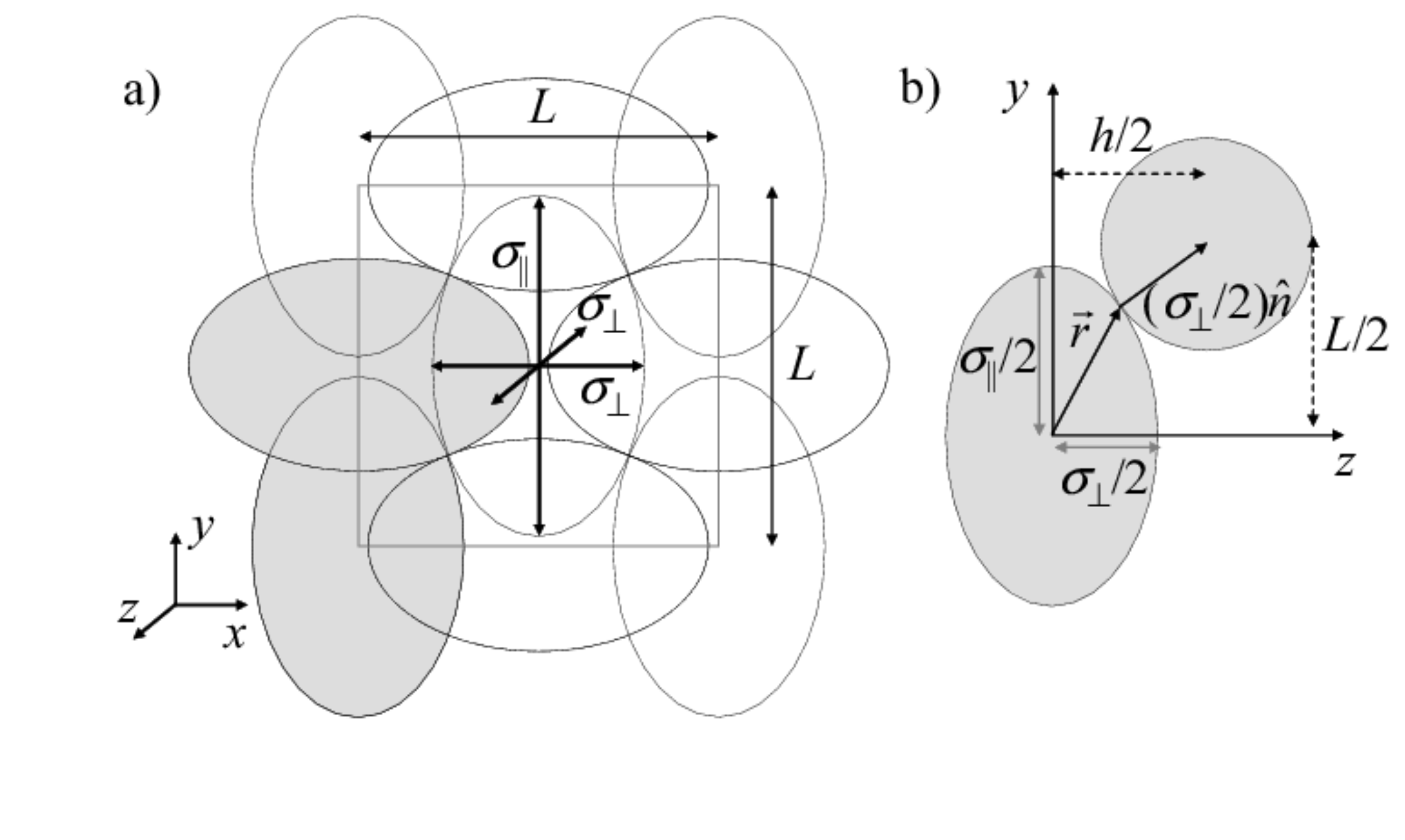}} 
\caption{\label{scheme} a) The figure illustrates the two-folded structure of the sm2 structure. b) Calculation of the distance between two equally oriented layers, $h$, for prolates (oblates result is identical).} 
\end{figure} 

Donev et.~al.~\cite{Donev04a} described an arrangement of ellipsoids able to surpass the maximal density of the sfcc structure. This structure is generated starting from a horizontal layer (A) of equally oriented ellipsoids, such that 5 ellipsoids are fitted in contact inside a face-centered square lattice of side $L$ (see Fig.~\ref{scheme}a)). The two axes parallel to the horizontal plane are given by $\sigma_{\|}$ and $\sigma_{\bot}$, respectively. As we are assuming ellipsoids of revolution, the length of the axis perpendicular to this plane will be also given by $\sigma_{\bot}$. Within these conditions, the length of the square lattice is
\begin{equation}
\label{calcL}
L = \frac{2\alpha}{\sqrt{1+\alpha^2}}\sigma_{\bot}=\frac{2\delta}{\sqrt{1+\delta^2}}\sigma_{min}.
\end{equation}
Then, a second horizontal layer (B) is built on top of the first one following the same procedure, but rotated by $\pi/2$ around the perpendicular direction. The second layer is also shifted in such a way that the centers of the ellipsoids are located in the holes formed by the first layer, each ellipsoid of the second layer touching four ellipsoids of the first one. This procedure is repeated successively, leading to a stratification in the form ABABAB..., where each layer perfectly fits in the holes of the other. In both cases (prolates and oblates), the obtained structure is denser than the sfcc structure for $\delta \leq \sqrt{3}$, and leads to a density maximum when the ellipsoids in the face-centered layers touch six rather than four in-plane neighbors, which occurs for $\delta=\sqrt{3}$~\cite{Donev04a}. Thus, for $\delta \leq \sqrt{3}$, we are taking this structure as the one with the highest possible density.

The distance between two equally oriented layers, $h$, can be analytically determined by considering that the ellipsoids of two consecutive layers are in contact. This calculation can be simplified to a two-dimensional problem in the vertical plane ($yz$-plane of Fig.~\ref{scheme}): We only need to determine the contact point between an ellipse of axes $(\sigma_{\bot},\sigma_{\|})$ and a circumference of diameter $\sigma_{\bot}$. Fig.~\ref{scheme}b) illustrates the position and orientation of both geometrical objects, which correspond to the shaded ellipsoids of Fig.~\ref{scheme}a). Setting the origin of the Cartesian coordinates at the center of the ellipse, the set of points that belong to the ellipse can be written as $\vec{r}=(y,z)=(1/2)(\sigma_{\|}\cos \phi,\sigma_{\bot}\sin \phi)$. With this parameterization, the unit vector perpendicular to the ellipse at the contact point may be expressed as $\hat{n}=(\sigma_{\bot}\cos \phi,\sigma_{\|}\sin \phi)/(\sigma_{\bot}^2\cos^2\phi+\sigma_{\|}^2\sin^2\phi)^{1/
2}$. The vector joining the centers of the ellipse and the circumference is
\begin{equation}
\vec{r}+(\sigma_{\bot}/2)\hat{n}=(h/2,L/2).
\end{equation}
This leads to the following set of equations
\begin{equation}
\label{Lv}
(L/\sigma_{\bot}-\alpha v)^2[(1-\alpha^2)v^2+\alpha^2]=v^2
\end{equation} 
\begin{equation}
\label{hv}
h/\sigma_{\bot}=\left(1-\alpha^2+\frac{\alpha}{v}\frac{L}{\sigma_{\bot}}\right)\sqrt{1-v^2},
\end{equation} 
where $v=\cos \phi$. Eq.~\ref{Lv} is a quartic equation that must be solved together with Eq.~\ref{calcL}. It has only one solution for $v$ in the range $[0,1]$. Once $v$ is known, $h$ is easily obtained from Eq.~\ref{hv} as a function of the aspect ratio, $\alpha$.

\begin{figure}
\resizebox{0.47\textwidth}{!}{\includegraphics{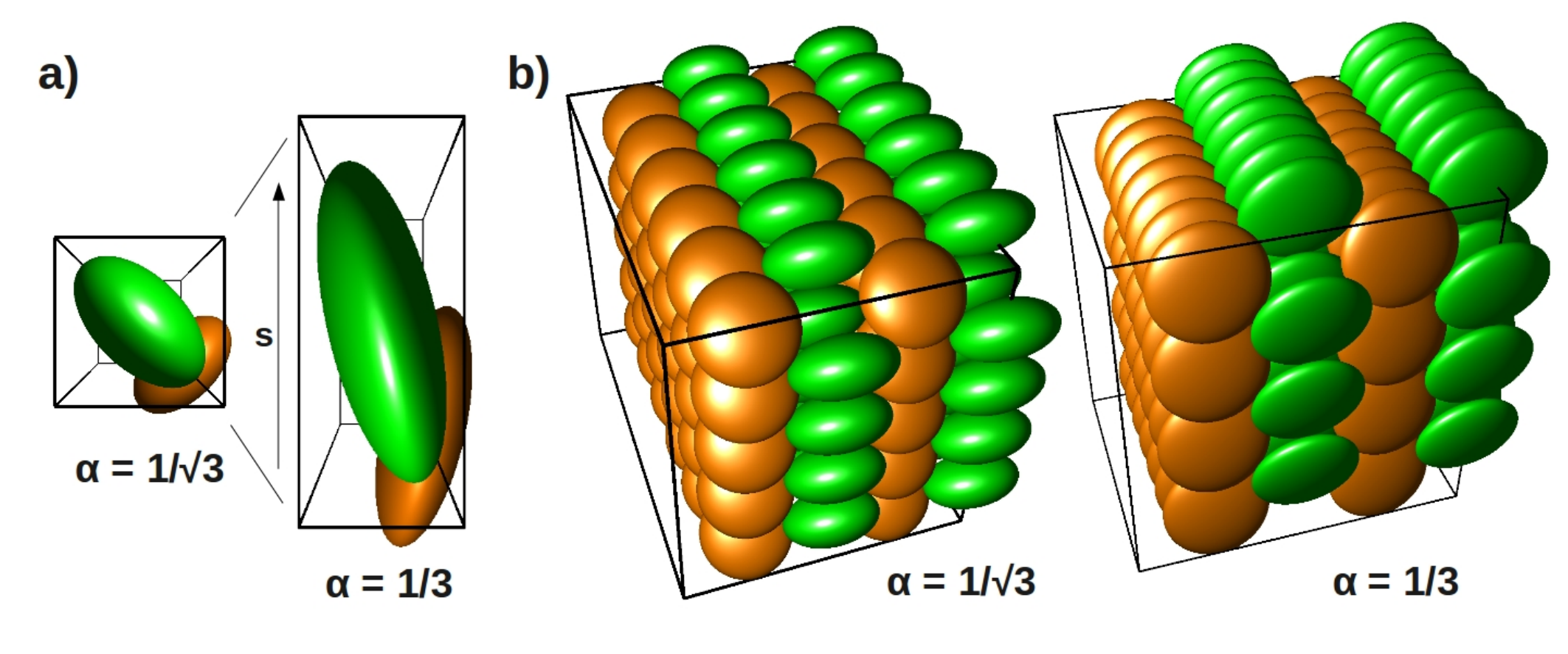}}
\caption{\label{celdas} a) Perspective views of unit cells for oblates with $\alpha = 1/\sqrt{3}$ (left), and the one obtained after stretching the particles and the unit cell (right), using a scale factor of $s=2.484$, to produce oblates with $\sigma_{\|}=1.198$ and $\sigma_{\bot}=3.594$ ($\alpha = 1/3$). b) Initial configurations for 120 oblates with $\alpha = 1/\sqrt{3}$ (left) and $\alpha = 1/3$ (right). Different colors are used to highlight the two different orientations.}
\end{figure}

The unit cell for this case is shown on the left of Fig.~\ref{celdas}a). It is a prism with square base of side $L/\sqrt{2}$ and height $h$, which contains a couple of particles with one of their principal axes (not the axis of revolution) along the prism height and the other two parallel to the prism base. One particle is placed at one (any) of the prism vertices with its axes at the base plane forming a $\pi/4$ angle with any given side of the base. The other particle is placed at the prism center, with both principal axes parallel to the prism base rotated $\pi/2$ with respect to the particle at the vertex. These two particles generate the two-folded structure when replicating the cell, one yielding layer A and the other layer B, both parallel to the prism base. This is shown on the left of Fig.~\ref{celdas} b) for oblates with $\delta= \sqrt{3}$. Then, the volume fraction is
\begin{equation}
\varphi = \frac{2v_{e}}{(L/\sqrt{2})^2h}=\frac{2\pi\alpha}{3(L/\sigma_{\bot})^2(h/\sigma_{\bot})},
\end{equation}
where we used that the volume of the ellipsoid is $v_{e}=\pi \sigma_{\bot}^2 \sigma_{\|}/6$. This equation predicts a continuous growth of the packing fraction as we increase the asymmetry from the value for spheres ($\delta =1$), until it finally reaches a saturation value for $\delta=\sqrt{3}$, where each ellipsoid touches six in-plane ellipsoids instead of four. The packing fraction obtained at this saturation point is $\varphi_{m}=0.77073$~\cite{Donev04a}.

For $\delta>\sqrt{3}$, Donev et.~al.~noticed that it is possible to stretch the structure obtained for $\delta=\sqrt{3}$, increasing $\delta$ by the same factor for ellipsoids belonging to different layers. Hence, this should lead to any desired aspect ratio $\delta>\sqrt{3}$ while preserving a mono-component system and keeping the same maximum density. The stretching can be done in a direction perpendicular to the layers (in general, this would not lead to ellipsoids of revolution), or in a direction parallel to any diagonal of the face-centered square lattice of layers A or B, since they coincide. In the case of our unit cell, this would be in the direction of any of the base sides. The relationship between the stretching factor, $s$, and $\delta$ is given by~\cite{Donev04a} 
\begin{equation}\label{stretching}
\delta^2=\frac{(2+s^2+2s^4)+ 2(1+s^2) \sqrt{1-s^2+s^4}}{3s^2}.
\end{equation}
Note that the stretching leads to an increase of both, $\sigma_{max}$ and $\sigma_{min}$. After the stretching, the new (elongated) axes, $\sigma_{max}'$ and $\sigma_{min}'$, are now
\begin{equation}\label{stretching-a2}
\sigma_{max}'= \sqrt{\frac{2 \delta^2 (1+s^2)}{(1+\delta^2)}}\sigma_{min}   \ \ , \ \ \sigma_{min}'=\sigma_{max}'/\delta.
\end{equation}
The sides of the prism become $L/\sqrt{2}$, $sL/\sqrt{2}$, and $h$, so that its base is now a rectangle of sides $L/\sqrt{2}$ and $sL/\sqrt{2}$. In this case $h$ can be easily obtained from $h=4v_{e}/(\varphi_m sL^2)$. The center of both particles are kept in the prism vertex and center, and their angles between the largest principal axis and the direction parallel to the prism side of length $L/\sqrt{2}$ turns into
\begin{equation}\label{stretching-alpha}
\psi=1/2\left(\pi \pm \arctan \left(\frac{s}{s^2-1}\right)\right). 
\end{equation}
These angles are $\pi/4$ (minus sign) and $3\pi/4$ (plus sign) for $s=1$ (no stretching), and $\pi/2$ for $s\rightarrow \infty$. The stretched cell is shown on the right of Fig.~\ref{celdas} a). In the simulations, we set $\sigma_{min}=1$, so the cell and particles are rescaled by $1/\sigma_{min}'$. A simulation snapshot of 120 oblates with $\delta = 3$ is shown on the right of Fig.~\ref{celdas} b). The cells before and after the stretching are particular cases of the more general sm2 family of structures~\cite{Pfleiderer07,Radu09}.
   
\subsection{Replica Exchange Monte Carlo}
\label{section-REMC}
This technique was developed to enhance sampling at difficult (high density / low temperature) conditions~\cite{Marinari92,Lyubartsev92,hukushima96}. It is based on the definition of an extended ensemble whose partition function is given by $Q_{extended}=\prod_{i=1}^{n_r}Q_{i}$, being $n_r$ the number of ensembles and $Q_i$ the partition function of ensemble $i$. This extended ensemble is sampled by $n_r$ replicas, each replica placed at each ensemble. The extended ensemble justifies the introduction of swap trial moves between any two replicas, whenever the detailed balance condition is satisfied. For hard particles, it is convenient to make use of isobaric-isothermal ensembles and perform the ensemble expansion in pressure~\cite{Odriozola09}. For this particular choice, the partition function of the extended ensemble turns~\cite{Okabe01,Odriozola09} 
\begin{equation}
Q_{\rm extended}=\prod_{i=1}^{n_r} Q_{N T P_i},
\end{equation} 
where $Q_{NTP_i}$ is the partition function of the isobaric-isothermal ensemble of the system at pressure $P_i$, temperature $T$, and with $N$ particles. 

We implemented a standard sampling of the $NTP_i$ ensembles, involving independent trial displacements, rotations of single ellipsoids, and volume changes. To increase the degrees of freedom of our small systems ($N=120$), we implemented non-orthogonal parallelepiped cells. Thus, sampling also includes trial changes of the angles and relative length sides of the lattice vectors defining the simulation cell. This is done while rescaling the cell sides and particles positions to preserve volume and keep a simple acceptation rule. Swap moves are performed by setting equal probabilities for choosing any adjacent pairs of replicas, and using the following acceptance rule~\cite{Odriozola09}
\begin{equation}
\label{accP} 
P_{\rm acc}\!=\! \min(1,\exp[\beta(P_i-P_j)(V_i-V_j)]), 
\end{equation} 
where $\beta=1/(k_BT)$ is the reciprocal temperature and $V_i-V_j$ is the volume difference between replicas $i$ and $j$. Adjacent pressures should be close enough to provide reasonable swap acceptance rates between neighboring ensembles. In order to take good advantage of the method, the ensemble at the smaller pressure must also ensure large jumps in configuration space, so that the higher pressure ensembles can be efficiently sampled.

Simulations started from the Donev-sm2 structures described in the previous section. We first perform about $5 \times 10^{12}$ trial moves at the desired state points, during which we check that the replicas have reached a stationary state (thermalization stage). It should be noted that achieving a stationary state requires considerably less simulation steps when starting from these ordered configurations than when starting from loose random configuration cells~\cite{Odriozola12}. We then perform $1 \times 10^{13}$ additional sampling trials. Maximum particle displacements, maximum rotational displacements, maximum volume changes, and maximum changes of the lattice vectors are adjusted for each pressure to yield acceptance rates close to 0.3. Since an optimal allocation of the replicas should lead to a constant swap acceptance rate for all pairs of adjacent ensembles~\cite{Rathore05}, we implemented a simple algorithm to smoothly adjust the intermediate pressures while keeping the maximum and minimum 
pressures fixed. To start the simulations, we use a geometric progression of the pressure with the replica index. These adjusting procedures are performed only during the equilibrating stage. Verlet neighbor lists~\cite{Donev05a,Donev05b} are used to improve performance. 
We set $N=120$ ellipsoids and $n_r=64$ (to cover a wide range of densities while keeping large swap acceptance rates). More details are given in previous works~\cite{Odriozola12,GuevaraHE}.

\section{Results}

\subsection{Hard ellipsoids phase diagram}

\begin{figure}
\resizebox{0.45\textwidth}{!}{\includegraphics{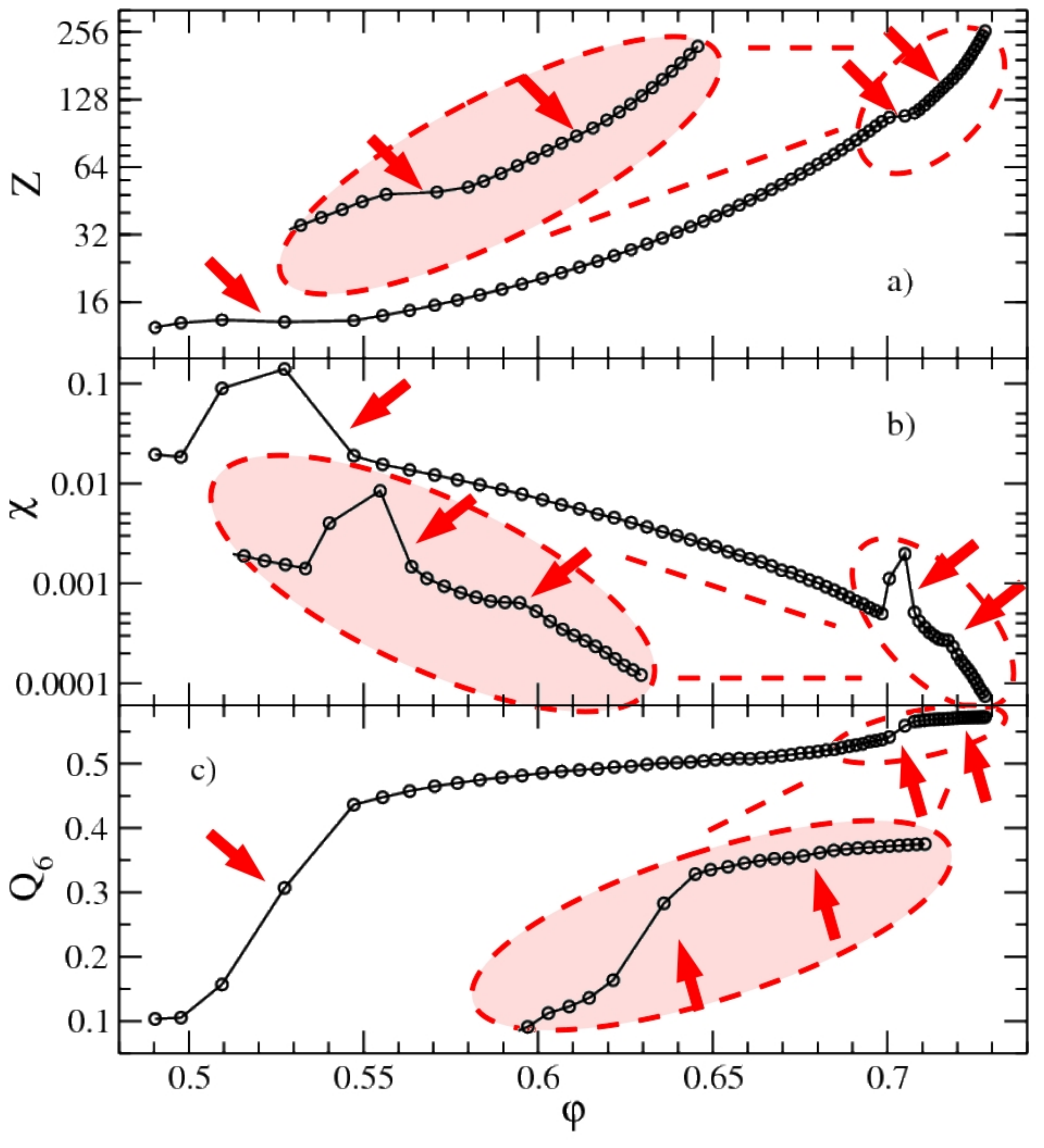}}
\caption{\label{ob-1.2-ZXQ}  a) Equation of state for oblates with $\delta = 1.2$, $Z(\varphi)$. b) Isothermal compressibility, $\chi(\varphi)$. c) Order parameter, $Q_6(\varphi)$. The insets zoom in the highlighted regions. Arrows point out the fluid-plastic, plastic-sfcc, and sfcc-sm2 phase transitions.}
\end{figure}

In the initial stage, all simulation replicas start with a sm2 structure. Then, we let the system of replicas decompress to yield a stationary state. Once this final state is reached, the compressibility factor ($Z(\varphi)=\beta P/\rho$), the isothermal compressibility ($\chi(\varphi)=N(<\rho^2>-<\rho>^2)/<\rho>^2$) and the $Q_6$-order parameter ($Q_6(\varphi)=\left(\frac{4\pi}{13}\sum_{m=-6}^{m=6}|<\!Y_{6m}(\theta,\phi)\!>|^2\right)^{1/2}$) are calculated. In these expressions $\rho$ is the particle number density and $<\!Y_{6m}(\theta, \phi)\!>$ is the average over all bonds and configurations of the spherical harmonics of the orientation polar angles $\theta$ and $\phi$~\cite{Steinhardt83,Rintoul96b,TorquatoRev}. Fig.~\ref{ob-1.2-ZXQ} shows all these quantities for quasi-spherical oblates ($\delta = 1.2$). The graphs also include arrows pointing to the phase transitions. Moving from low to high pressures, we first find a fluid-solid phase transition, where an isotropic fluid and a plastic crystal 
coexist~\cite{Frenkel85}. Then, there is a solid-solid transition between a plastic crystal and a sfcc crystal. Finally, at high density we observe another solid-solid transition, between the sfcc structure and the densest sm2 crystal. The first two transitions are first order, as pointed out by the $Z(\varphi)$ discontinuity (see Fig.~\ref{ob-1.2-ZXQ} a)) and by the formation of bimodal density distributions (not shown). The sfcc-sm2 transition is not first order, according to the continuous $Z(\varphi)$ (Fig.~\ref{ob-1.2-ZXQ} a)), the very small kink of $\chi(\varphi)$ (Fig.~\ref{ob-1.2-ZXQ} b)), and the slight deformation of the Gaussian density distributions (not shown). At this point we must stress that this last conclusion may be affected by the small system sizes we are considering. The order parameter $Q_6(\varphi)$ also provides evidences of the transitions, as shown by the arrows in panel c). Here, a very subtle increase of $Q_6(\varphi)$ is observed for the sfcc-sm2 transition. 

\begin{figure}
\resizebox{0.4\textwidth}{!}{\includegraphics{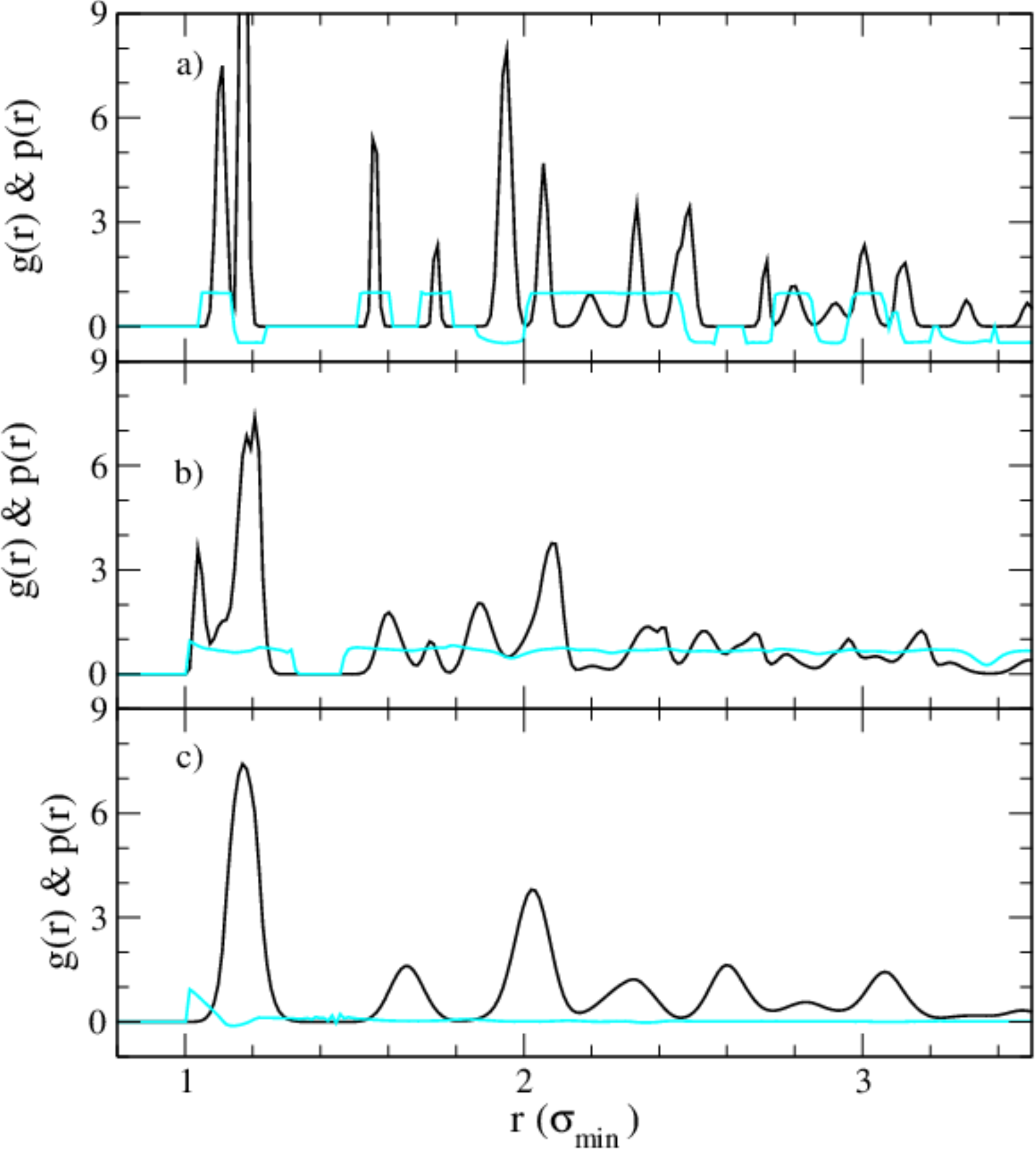}}
\caption{\label{ob-1.2-gdr} Radial distribution functions, $g(r)$, (dark) and their corresponding radial orientational order parameter functions, $p(r)$, (light gray) for $\delta = 1.2$, and for $Z=259$, 111, and 60, as shown in panels a), b) and c), respectively. The corresponding snapshots are shown in panels a) (sm2), b) (sfcc), and c) (plastic crystal) of Fig.~\ref{ob-1.2-snapshots}, respectively. }
\end{figure}

We can get a more clear evidence of the solid-solid phase transitions by studying the behavior of the radial distributions functions, $g(r)$, and the radial orientational order parameter functions, $p(r)$, defined as $p(r)=<(3(\hat{\mathbf{u}}_{i} \cdot \hat{\mathbf{u}}_{j})^2-1)>\!/2$~\cite{Eppenga1984}. They are shown in Fig.~\ref{ob-1.2-gdr}. Panels a), b), and c) of Fig.~\ref{ob-1.2-gdr} are built for a high ($Z=259$), intermediate ($Z=111$), and low pressure ($Z=60$) solids, respectively. Thence, $g(r)$ and $p(r)$ of panel a) are ensemble averages of, mostly, sm2 structures. For panel b) the dominant structure is a sfcc and for panel c) the most frequent is a plastic solid. The $g(r)$ for the sm2 structure shows two main peaks, a first one at a distance close but larger than $\sigma_{min}$, and a second larger one, at a somewhat larger distance. These two peaks correspond to the first coordination shell, for the intra (four) and extra-plane (eight) neighbors, respectively (see Fig.~\ref{scheme}). As 
shown by the $p(r)$ function, the first $p(r)\simeq 1$ and the second $p(r)\simeq -0.5$ peaks correspond to the parallel (intra-plane) and perpendicular (extra-plane) orientations, respectively. Thus, in general for this arrangement, a positive $p(r)$ is associated with $g(r)$ peaks for particles belonging to the same plane (A or B) (see section~\ref{section-Donev}), whereas a negative $p(r)$ corresponds to correlations between particles of planes A and B. Contrasting with the $p(r)$ function of Fig.~\ref{ob-1.2-gdr} a), panel b) shows a positive $p(r)$ for all distances, $r$. This implies a background parallel long-range orientational particle-particle correlation. Furthermore, the distance between the first and second $g(r)$ peaks enlarges, pointing out the differentiation between the stretched and the unstretched sides of the sfcc cell. Finally, panel c) shows the $g(r)$ and $p(r)$ functions of a plastic crystal. That is, while the $g(r)$ still shows the well-defined peaks of a solid (fcc-like), $p(r)$ 
shows no 
long-range angular correlations.       

\begin{figure}
\resizebox{0.47\textwidth}{!}{\includegraphics{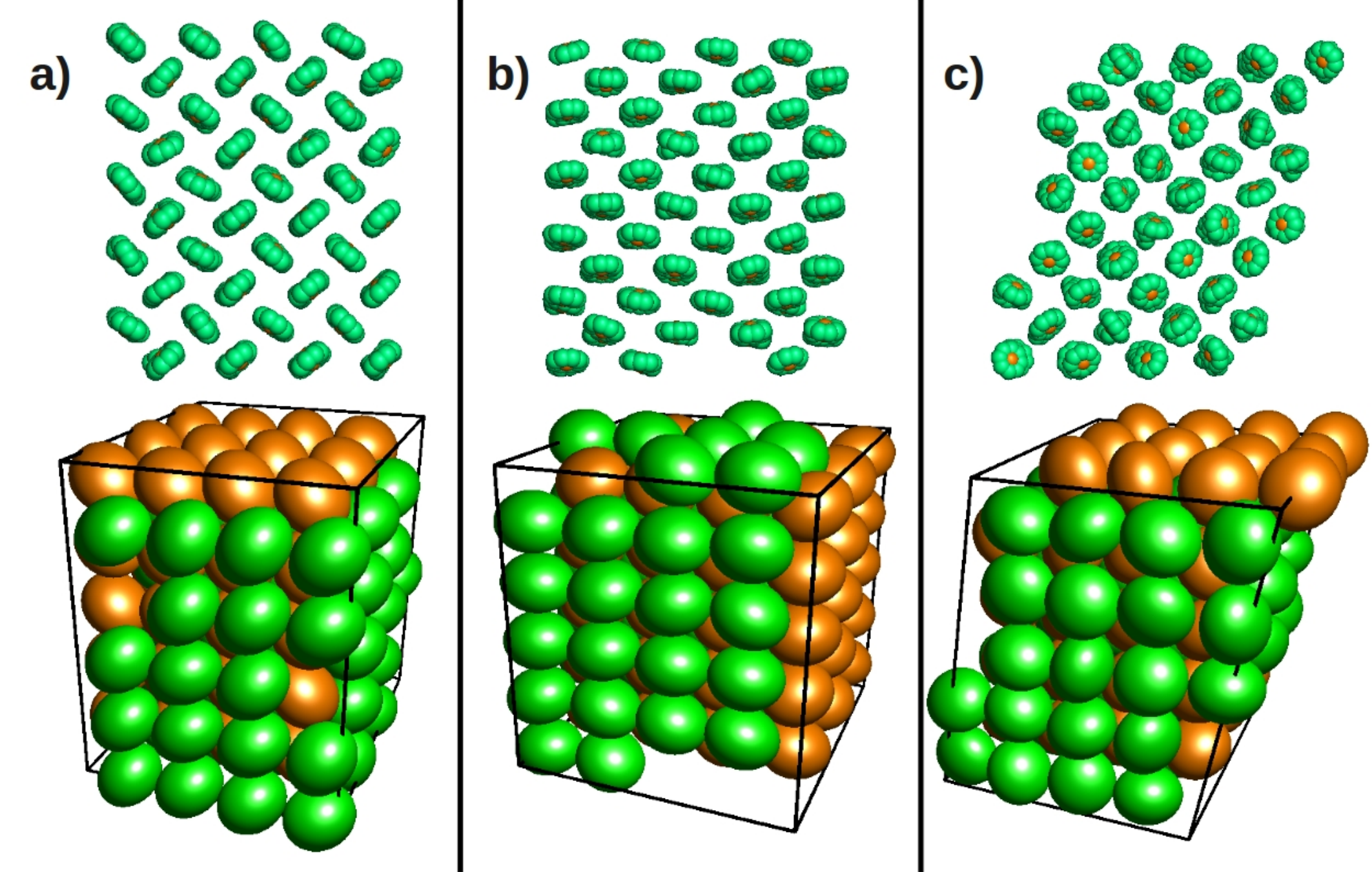}}
\caption{\label{ob-1.2-snapshots} Equilibrium structures for oblates with $\delta = 1.2$. Lower panels show snapshots where ellipsoids belonging to the same layer are equally colored. Upper panels show the corresponding front views where oblates are represented by small plate-like particles to highlight their crystal-like positions and orientations. Columns a), b), and c) correspond to sm2, sfcc, and plastic solids. Pressure decreases from a) to c).}
\end{figure}

All features of the $g(r)$ and $p(r)$ functions of Fig.~\ref{ob-1.2-gdr} correspond to the structures shown in Fig.~\ref{ob-1.2-snapshots}. The lower panels of Fig.~\ref{ob-1.2-snapshots} illustrate snapshots of the sm2, sfcc, and plastic solids, in correspondence with panels a), b), and c) of Fig.~\ref{ob-1.2-gdr}. To gain clarity, the upper panels of this figure show the corresponding front views where the oblates are represented by small plate-like particles. Also, particles in the lower panels are colored according to their positions, to highlight the layering of the structures. From this pictures, the sm2, sffc, and plastic-solid structures can be easily recognized. 

\begin{figure}
\resizebox{0.46\textwidth}{!}{\includegraphics{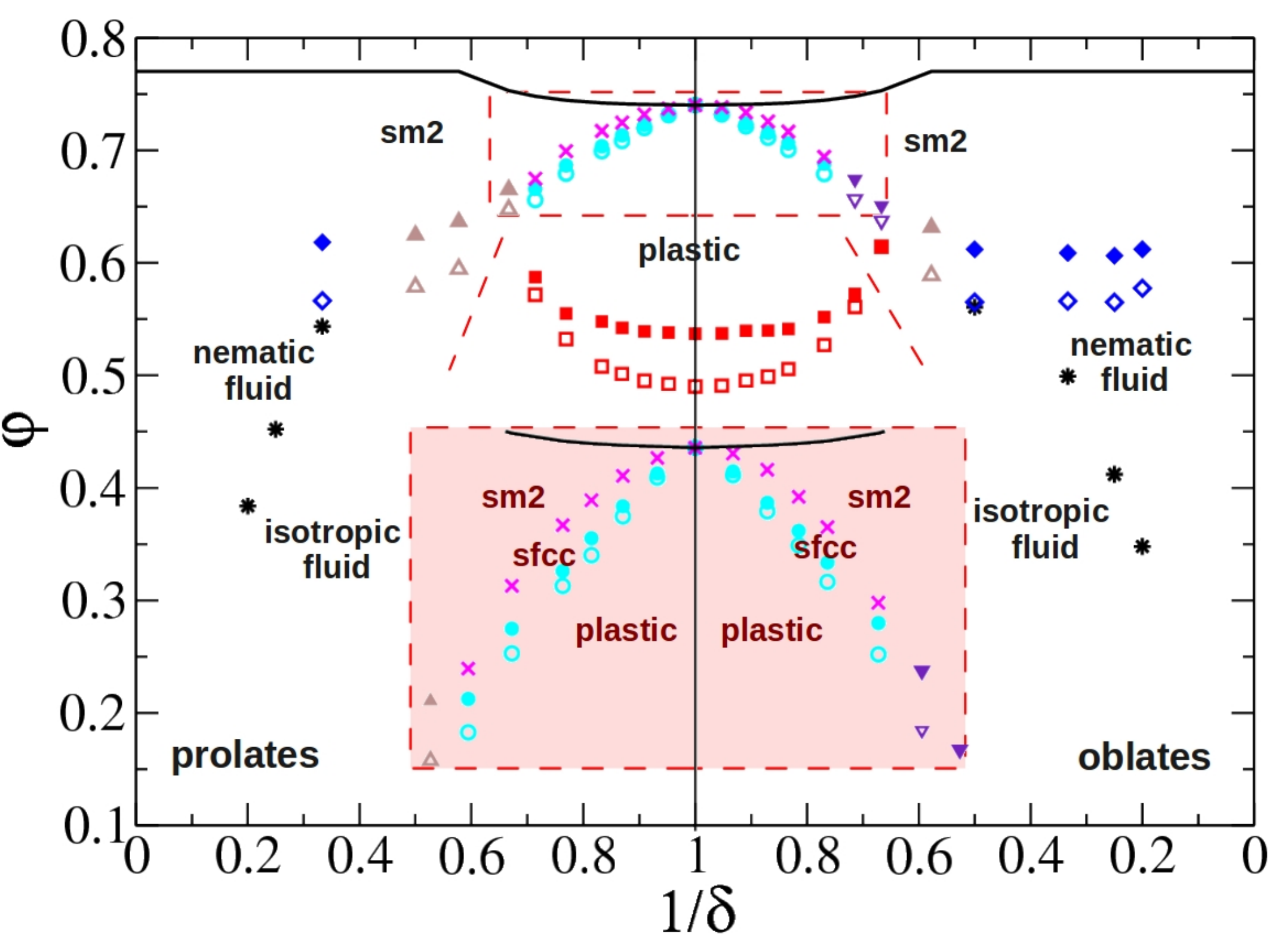}}
\caption{\label{phase-diagram} Phase diagram of hard ellipsoids of revolution. The spherical case is given for $\delta=1$, whereas prolates are at the left and oblates at the right. The dark solid line is the maximally achievable density~\cite{Donev04a}. There are several transition types. These are: Isotropic-nematic fluid-fluid (asterisks), isotropic-plastic fluid-solid (squares), nematic-sm2 fluid-solid (diamonds), isotropic-sm2 fluid-solid (upward triangles), plastic-sm2 solid-solid (downward triangles), plastic-sfcc solid-solid (circles), and sfcc-sm2 solid-solid (crosses). Pairs of open and solid symbols are employed to show coexistence regions. Single symbols are employed to point out higher order transitions. The inset zooms in the sfcc stable region (in-between solid circles and crosses). }
\end{figure}

\begingroup
\squeezetable
\begin{table*}
\caption{Coexistence volume fraction borders $\varphi$ and pressure $P^*=\beta P \sigma_{min}^3$ of transitions for cases with low asymmetry. Subindexes $l$ and $h$ refer to the low and high density borders. A dash means the absence of the transition. Errors are always below $3\%$ for all quantities.} \label{Table1}
\begin{tabular}{||c|c c c|c c c|c c c|c c||}
\hline
\hline
 & \multicolumn{3}{c|}{isotropic-plastic} & \multicolumn{3}{c|}{plastic-sm2} & \multicolumn{3}{c|}{plastic-sfcc} & \multicolumn{2}{c||} {sfcc-sm2} \\
$\alpha$ & $\varphi_l$ & $\varphi_h$ & $P^*$ & $\varphi_l$ & $\varphi_h$ & $P^*$ & $\varphi_l$ & $\varphi_h$ & $P^*$ & $\varphi$ & $P^*$\\
\hline
1.400 & 0.572 & 0.587 & 18.3 & - & - & - & 0.656 & 0.666 & 47.4 & 0.675 & 48.9 \\
1.300 & 0.532 & 0.555 & 12.9 & - & - & - & 0.679 & 0.687 & 69.7 & 0.699 & 85.6 \\
1.200 & 0.508 & 0.548 & 11.5 & - & - & - & 0.699 & 0.704 & 116 & 0.717 & 172 \\
1.150 & 0.501 & 0.542 & 10.9 & - & - & - & 0.708 & 0.713 & 156 & 0.725 & 269 \\
1.100 & 0.495 & 0.539 & 10.8 & - & - & - & 0.720 & 0.723 & 265 & 0.732 & 535 \\
1.050 & 0.492 & 0.538 & 10.8 & - & - & - & 0.731 & 0.733 & 634 & 0.737 & 1528 \\
1.000 & 0.490 & 0.537 & 10.5 & - & - & - & 0.740 & 0.740 & $\infty$ & 0.740 & $\infty$ \\
0.952 & 0.491 & 0.537 & 10.3 & - & - & - & 0.732 & 0.733 & 661 & 0.739 & 2412 \\
0.909 & 0.495 & 0.540 & 9.91 & - & - & - & 0.721 & 0.724 & 256 & 0.734 & 605 \\
0.870 & 0.499 & 0.540 & 9.40 & - & - & - & 0.711 & 0.716 & 152 & 0.726 & 251 \\
0.833 & 0.506 & 0.541 & 9.08 & - & - & - & 0.701 & 0.706 & 100 & 0.717 & 141 \\
0.769 & 0.527 & 0.552 & 9.48 & - & - & - & 0.679 & 0.688 & 54.3 & 0.694 & 59.0 \\
0.714 & 0.561 & 0.572 & 11.2 &  0.657 & 0.674 & 33.1 & - & - & - & - & - \\
0.667 & 0.614 & 0.614 & 16.3 &  0.638 & 0.651 & 20.2 & - & - & - & - & - \\
\hline
\hline
\end{tabular}
\end{table*}
\endgroup

\begingroup
\squeezetable
\begin{table*}
\caption{Coexistence volume fraction borders $\varphi$ and pressure $P^*=\beta P \sigma_{min}^3$ of transitions for cases with large asymmetry. Subindexes $l$ and $h$ refer to the low and high density borders. A dash means the absence of the transition whereas a blank means that the experiment was not carried out. Errors are always below $4\%$ for all quantities.} \label{Table2}
\begin{tabular}{||c|c c|c c c|c c c||}
\hline
\hline
 & \multicolumn{2}{c|}{isotropic-nematic} & \multicolumn{3}{c|}{isotropic-sm2} & \multicolumn{3}{c||}{nematic-sm2} \\
$\alpha$ & $\varphi$ & $P^*$ & $\varphi_l$ & $\varphi_h$ & $P^*$ & $\varphi_l$ & $\varphi_h$ & $P^*$\\
\hline
5.000 & 0.384 & 1.57 & - & - & - &  &  &  \\
4.000 & 0.452 & 3.14 & - & - & - &  &  &  \\
3.000 & 0.543 & 8.30 & - & - & - & 0.566 & 0.618 & 9.50\\
2.000 & - & - & 0.579 & 0.625 & 15.3 & - & - & - \\
1.732 & - & - & 0.595 & 0.637 & 19.3 & - & - & - \\
1.500 & - & - & 0.648 & 0.665 & 35.4 & - & - & - \\
0.577 & - & - & 0.589 & 0.632 & 10.5 & - & - & - \\
0.500 & 0.561 & 6.52 & - & - & - & 0.565 & 0.612 & 6.91 \\
0.333 & 0.499 & 1.87 & - & - & - & 0.566 & 0.609 & 2.99 \\
0.250 & 0.412 & 0.550 & - & - & - & 0.565 & 0.606 & 1.61 \\
0.200 & 0.348 & 0.229 & - & - & - & 0.577 & 0.612 & 1.09 \\
\hline
\hline
\end{tabular}
\end{table*}
\endgroup

The existence of a subtle sfcc-sm2 transition for oblates with $\delta=1.2$ encouraged us to explore the boundaries of the sfcc stable region. For this purpose, we performed a similar analysis, but now considering other aspect radios for both oblate and prolate cases. The obtained results allow the construction of a refreshed hard-ellipsoid phase diagram, which is shown in Fig.~\ref{phase-diagram}. The diagram is split in half by the hard sphere case ($\delta=1$). Prolate cases are at the left and oblate systems are at the right of this vertical line. Particles' asymmetry increases by moving away from this central line. The extreme cases are infinitely narrow needles ($1/\delta \rightarrow 0$ at the left) and infinitely thin plates ($1/\delta \rightarrow 0$ at the right). Prolates with $\delta>3$ are not included since a larger number of particles is needed to fulfill the minimum-image convention for all densities. At high densities, we are placing the (currently accepted) maximally achievable density~\cite{Donev04a} as a black solid line. All transitions found in our simulations are indicated in this chart. Isotropic-nematic fluid-fluid transitions are given as asterisks, isotropic-plastic fluid-solid transitions as squares, nematic-sm2 fluid-solid transitions as diamonds, isotropic-sm2 fluid-solid transitions as upward triangles, plastic-sm2 solid-solid transitions as downward triangles, plastic-sfcc solid-solid transitions as circles, and sfcc-sm2 solid-solid transitions as crosses. Isotropic-nematic fluid-fluid and sfcc-sm2 solid-solid transitions are higher order, and thus, they are shown as a single point, located at the packing fraction where the isothermal compressibility reaches a local peak. All other transitions are first order, and thus, a couple of symbols are used to denote the borders of the coexistence regions. The phase boundaries are determined by means of the histogram re-weighting technique described elsewhere~\cite{Ferrenberg88,Ferrenberg89}. The numerical values are given in tables~\ref{Table1} and~\ref{Table2}. According to previous results, increasing the system size would slightly shift the transitions towards larger densities and would make them slightly wider~\cite{Odriozola09,GuevaraHE}. 

\begin{figure}
\resizebox{0.48\textwidth}{!}{\includegraphics{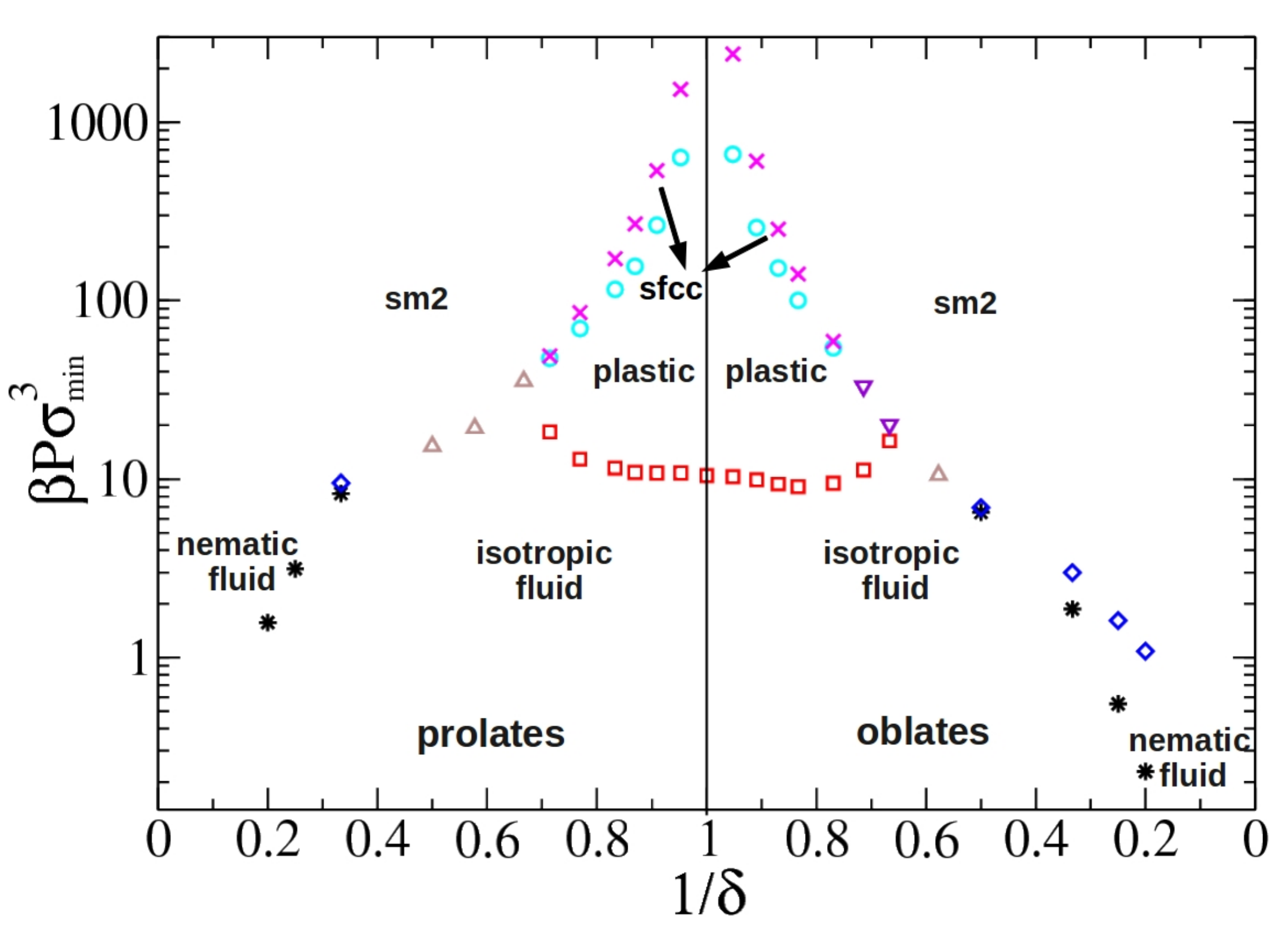}}
\caption{\label{pressure} Pressures at which transitions take place as a function of $\delta^{-1}$ for prolates (left) and oblates (right). Different symbols match the transitions given in Fig.~\ref{phase-diagram}. These are: Isotropic-nematic fluid-fluid (asterisks), isotropic-plastic fluid-solid (squares), nematic-sm2 fluid-solid (diamonds), isotropic-sm2 fluid-solid (upward triangles), plastic-sm2 solid-solid (downward triangles), plastic-sfcc solid-solid (circles), and sfcc-sm2 solid-solid (crosses). }
\end{figure}

The inset of Fig.~\ref{phase-diagram} illustrates a zoom of the high density area above the plastic region. There, it can be appreciated a narrow density region where the sfcc solid spontaneously forms. Hence, for very low asymmetries, the plastic solid turns into a sfcc solid before taking the form of a sm2 structure under very high compression. It should be noted that the sfcc solid region is not so small in terms of the pressure range at which it is stable. As shown in Fig.~\ref{pressure}, the sfcc stable region can be several hundreds of $\beta P \sigma^3_{min}$ wide (see also table~\ref{Table1}). The existence of a stable sfcc structure above the plastic solid region is in agreement with Radu et. al.~\cite{Radu09}. Nonetheless, and despite that the phase boundaries for the sfcc structure were not given, it was suggested that this structure should be the most stable for $1/\delta < 0.65$. Our data show that this is only true for a very small density region above the plastic solid, so a sfcc-sm2 
transition appears at very high pressures and densities. Free energy calculations through thermodynamic integration should confirm this finding~\cite{Frenkel19843188,Schilling2009,Romano2012}. As mentioned in the introduction, this transition should necessarily exist in order to be consistent with the fact that the sm2 structure is the one leading to the maximally achievable density for all aspect ratios.
  
\subsection{Comparing runs from random and ordered initial conditions}

\begin{figure}
\resizebox{0.4\textwidth}{!}{\includegraphics{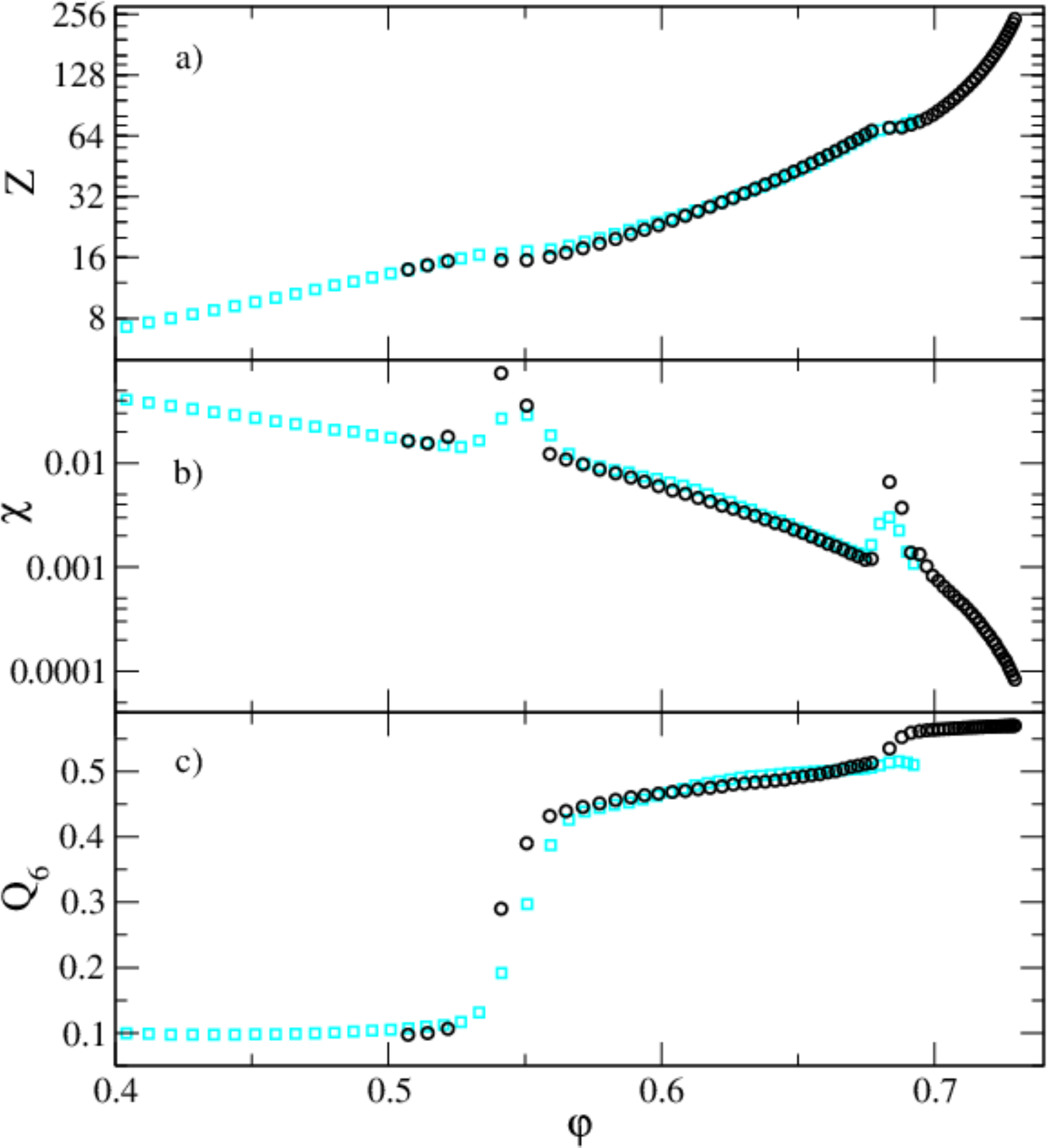}}
\caption{\label{oblates-1.3} Oblates with $\delta = 1.3$. a) Compressibility factor $Z(\varphi)$. b) Isothermal compressibility, $\chi(\varphi)$. c) Bond order parameter, $Q_6(\varphi)$. Dark circles correspond to simulations starting from dense sm2 structures, whereas light squares correspond to simulations starting from loose random cells (data taken from~\cite{Odriozola12}). }
\end{figure}

This section is devoted to compare the REMC results obtained by starting from loose random cells (data taken from~\cite{Odriozola12}) with those obtained from dense sm2 initial configurations. In order to support the ergodicity of the simulated systems, one should obtain the same results regardless of the initial configurations. For certain systems at very high pressures, the ergodic condition is difficult to achieve. This is simply because simulations tend to get stuck on configuration space at high densities. When this occurs, in turn, dynamical properties usually show solid-like behaviors. That is, particles relaxation times diverge, and diffusion coefficients turn practically zero~\cite{Berthier05}. Thus, with the aim of enhancing sampling at these difficult conditions, certain Monte Carlo techniques were developed, usually introducing unnatural displacements while keeping the detailed balance condition. This is the case of the REMC technique, where replicas can travel throughout different ensembles. 

As mentioned in section~\ref{section-REMC}, an improved sampling can be obtained whenever a relatively large swap acceptance rate is achieved between all set pressures. Nonetheless, this does not guaranty ergodicity. For instance, replicas may frequently jump from one side to the other of a coexistence region without residing during a large number of steps where they \textquotedblleft do not belong''. That is, the replica having the most compressed fluid-like configuration may swap positions with the replica having the less compressed solid-like structure, across a coexistence region, but this fluctuation may not last long enough to allow the fluid-like configuration become solid-like and vice-versa. Consequently, there would be a low rate of generation of new solid-like structures, and the improved sampling may be not sufficient to attain ergodic conditions. 

An example where REMC works well is given in Fig.~\ref{oblates-1.3}. There it is shown how the results obtained starting from dense sm2 structures (dark circle curves) match the ones started from loose random initial conditions (light square curves). This figure corresponds to oblates with $\delta = 1.3$, and considers the pressure range $2.0 <\beta P \sigma^3_{min}< 200$ for random initial conditions and $8.5 <\beta P \sigma^3_{min}< 1000$ for sm2 initial conditions. As observed, both runs present the same density jumps, pointing out the isotropic-plastic fluid-solid, and the plastic-sfcc solid-solid transitions. In fact, the dark $\chi(\varphi)$ curve also shows a sfcc-sm2 transition, appearing as a small shoulder developed at the right of the plastic-sfcc transition peak. This subtle transition, which is better seen from the slight distortion of the Gaussian density distributions and from the radial distribution functions (not shown), is not captured by the run with random initial conditions. Nevertheless,
 the fact that the plastic-sfcc solid-solid transition is captured by both runs is quite remarkable, given the extremely high density at which it occurs. In general, we observed good agreements between runs started from different conditions for ellipsoids with $\delta < 1.5$.  

\begin{figure}
\resizebox{0.4\textwidth}{!}{\includegraphics{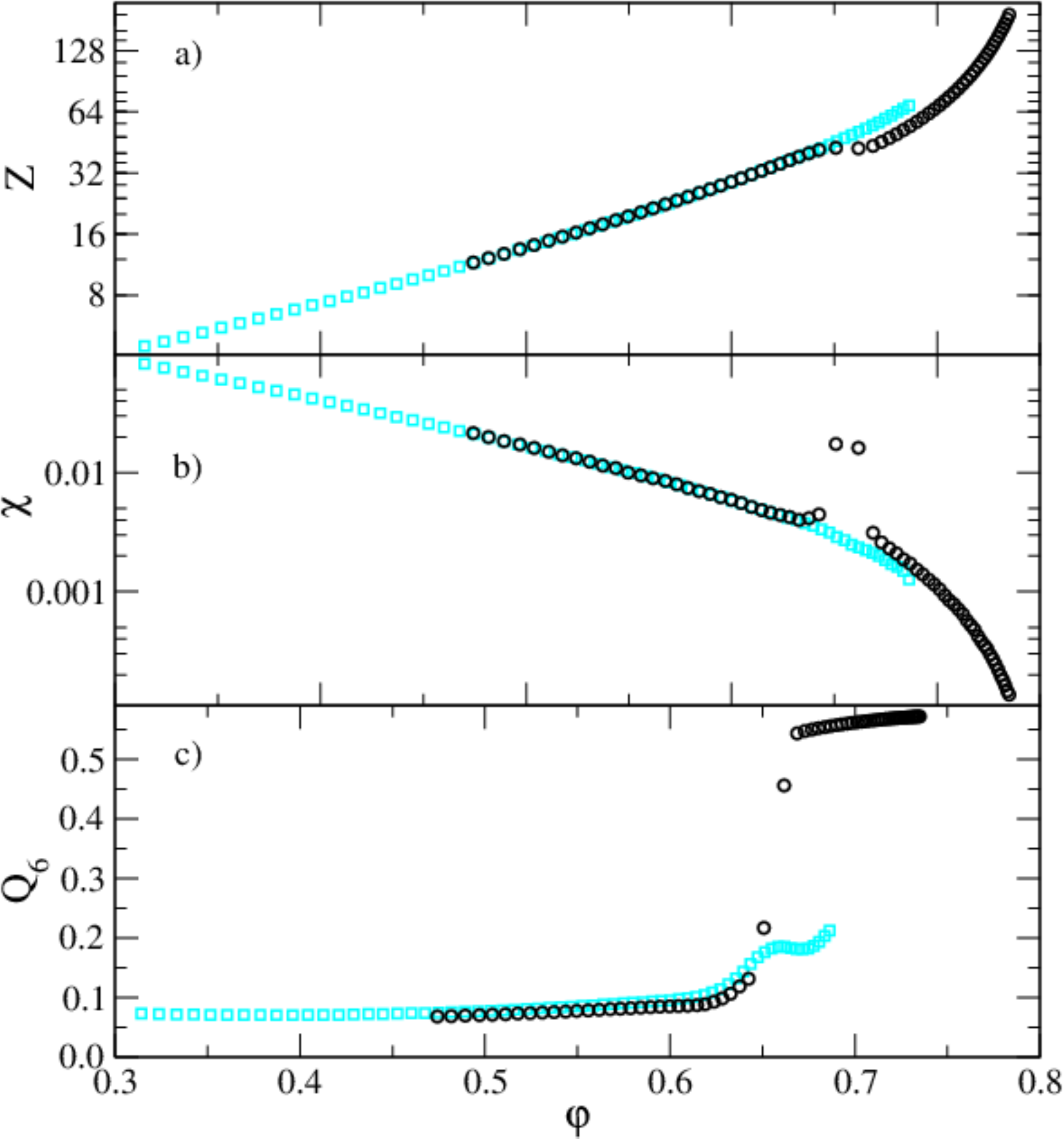}}
\caption{\label{prolates-1.5} Prolates with $\delta = 1.5$. a) Compressibility factor $Z(\varphi)$. b) Isothermal compressibility, $\chi(\varphi)$. c) Bond order parameter, $Q_6(\varphi)$. Dark circles correspond to simulations starting from dense sm2 structures, whereas light squares correspond to simulations starting from loose random cells (data taken from~\cite{Odriozola12}). }
\end{figure}

The comparison for prolates with $\delta = 1.5$ is shown in Fig.~\ref{prolates-1.5}. Here, it is seen a good match between the two cases only for densities below the isotropic-sm2 fluid-solid transition. In fact, the transition is not captured when starting from random initial configurations. Instead, the fluid curve extends towards higher densities, defining a metastable branch which cannot be broken even by a very large number of REMC cycles. This metastable branch is analogous but stronger than the one for hard spheres~\cite{Hermes10,PerezAngel11,Santos11}. It should be emphasized that the relationship $\chi = N  (\langle\rho^2\rangle-\langle\rho\rangle^2)/ \langle\rho\rangle^2 = \partial\rho/\partial(\beta P)$ is satisfied for both data sets. Fulfilling the above relationship, suggested to hold only at equilibrium~\cite{Santen00}, is then a necessary but not sufficient condition for equilibrium~\cite{Odriozola11}. Note that the order parameter $Q_6$ indicates the development of some bond order at the 
transition. Another curiosity is that an extrapolation of the metastable branch would intersect the solid branch. It should be said that for densities below but close to the transition, the isotropic fluid shows extremely low translational and rotational diffusion coefficients~\cite{DeMichele07,Pfleiderer08a,Pfleiderer08b}, i.~e. dynamics turns glassy. Thus, when starting from random initial configurations, the glassy dynamics not only hinders the solid formation when dynamics simulations are used, but also when REMC is the employed technique. The other way around, when REMC produces an apparent stationary state which differs from equilibrium at high densities, a glassy dynamics may be expected. The occurrence of disordered structures at very large densities was also found experimentally~\cite{Man05} and by means of computations~\cite{Donev07,TorquatoRev}. These references report that disordered states yield a maximally random jammed density, $\varphi \approx 0.712$, peaking at $\delta=1.5$. Additionally, 
they show that prolates produce slightly higher maximally random jammed densities than oblates.

\begin{figure}
\resizebox{0.4\textwidth}{!}{\includegraphics{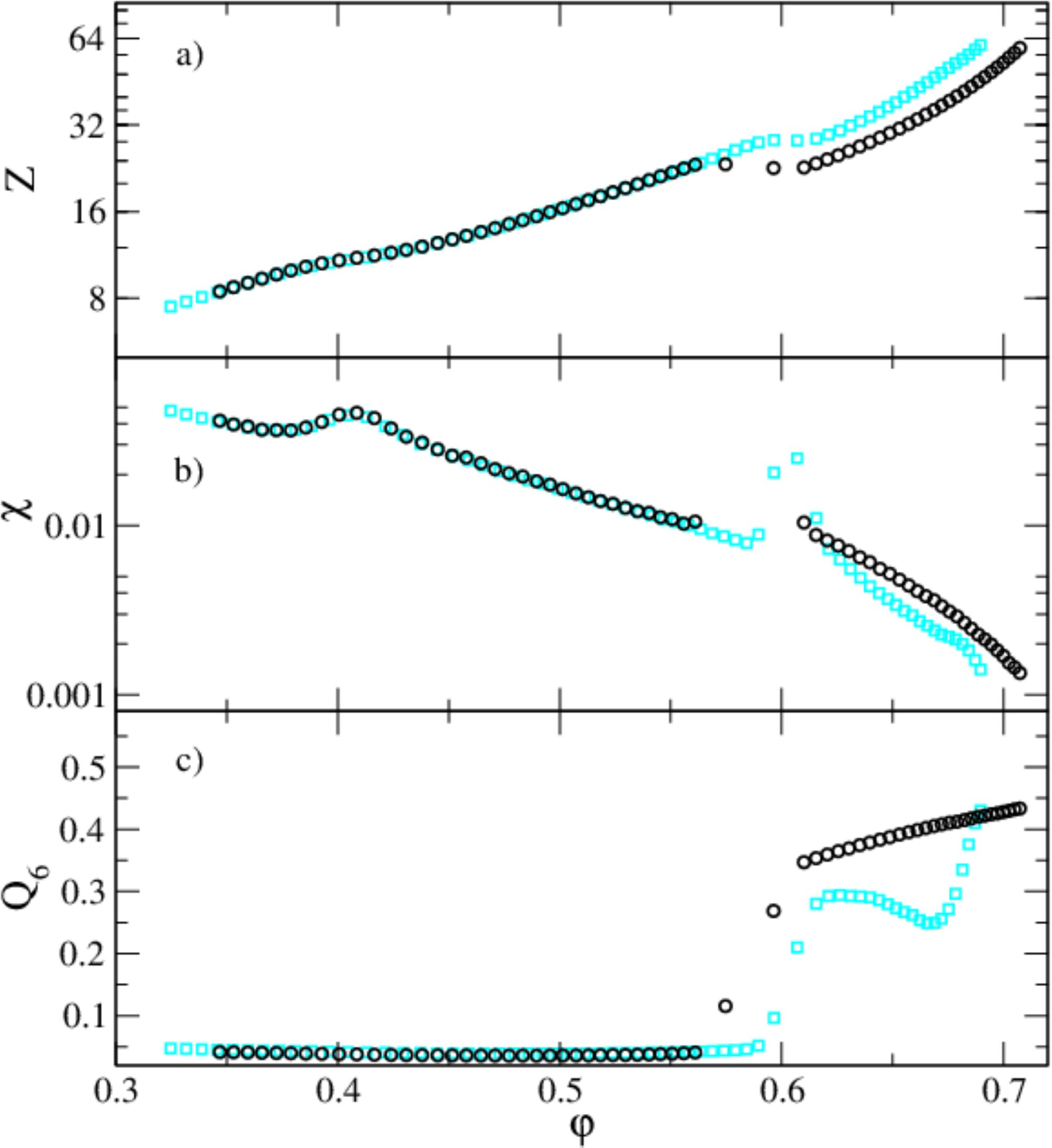}}
\caption{\label{prolates-4.0} Oblates with $\delta = 4$. a) Compressibility factor $Z(\varphi)$. b) Isothermal compressibility, $\chi(\varphi)$. c) Bond order parameter, $Q_6(\varphi)$. Dark circles correspond to simulations starting from dense sm2 structures, whereas light squares correspond to simulations starting from lose random cells (data taken from~\cite{Odriozola12}).}
\end{figure}

Another example where curves do not entirely agree is shown in Fig.~\ref{prolates-4.0}. Here again, a good match between the different runs is obtained for densities below the nematic-sm2 fluid-solid transition (the agreement on the description of the isotropic-nematic fluid-fluid transition is very good). For densities above the fluid-solid transition and for a given pressure, the run with initial random configurations yields smaller densities, isothermal compressibilities, and $Q_6$ values than the ones obtained starting from sm2 structures. Nonetheless, both runs capture a fluid-solid transition and, in addition, the resulting solid structures are similar. We then justify differences due to the appearance of imperfect sm2 structures when starting from disordered configurations. This turns evident from a snapshot overview (not shown). It should be mentioned that only certain box-shapes can hold an exact number of particles of a given perfect sm2 structure. This effect, for the system sizes we are employing,
 is important. In previous work~\cite{Odriozola12} we only let free the angles of the simulation cells. However, this procedure seems to be not enough to completely relax the solid phase. In the present work we are letting angles and sides to vary independently. Hence, systems with different degrees of freedom are being compared and this is why differences appear for the solid phases. 

Summing up, starting from random initial configurations has the following characteristics: a) It naturally produces highly dense structures which may assist proposing the equilibrium structure or confirm/refute the believed equilibrium structure. b) It allows the detection of strong metastable regions, which may be linked to systems showing extremely low dynamics. Thus, when these metastable regions appear for dense fluid-like structures, a glassy behavior can be expected. c) Long runs are frequently necessary to reach a stationary state. On the other hand, starting from a-priori set dense crystal structure has the following characteristics: a) When the crystal structure corresponds to equilibrium at high densities, the obtained high density branch should also correspond to equilibrium,  while transitions are easily captured by decompression. b) Thus, relatively short runs would produce equilibrium since metastable regions are avoided. c) The imposition of a crystal structure not representative of 
equilibrium may lead to a stationary state inside a strong metastable region (strategies to find structure candidates to reach maximally achievable densities, i.~e.~structure candidates for equilibrium at very high pressures, are given elsewhere~\cite{Torquato10,Graaf11,Torquato12}). From combining random and crystal initial conditions a better understanding of the system is possible, i.~e. by comparing results from both strategies one can detect metastable regions or support ergodicity.

\section{Conclusions}
Further details on the phase diagram of hard ellipsoids of revolution were captured by decompressing high density sm2 structures by means of replica exchange Monte Carlo simulations. Mainly, we observed a very rich behavior for quasi-spherical oblates and prolates. These systems, from low to high pressures, show the following phases: isotropic fluid, plastic solid, stretched-fcc solid, and sm2 solid. According to our data obtained for small system sizes, the first three transitions are first order, whereas the last one is a subtle, high order transition. This picture is consistent with the fact of having the sm2 structure capable of producing the maximally achievable density.  

Replica exchange Monte Carlo (REMC) simulations started from dense sm2 structures produce, by decompression of the initial configurations, equilibrium states for all set pressures. This, in addition, is yield in relatively few Monte Carlo steps. Thus, when there are structure candidates for the maximally achievable density, REMC simulations started by setting them as starting cells would provide a fast way to produce the whole phase diagram. 

\begin{acknowledgments}
G.B-C thanks CONACyT for a Ph.D.~scholarship and Dr.~Eliezer Braun Guitler for his support. A.M-J is grateful to the Ministerio de Econom\'{\i}a y Competitividad (Project No.~MAT2012-36270-C04-02), and the project CEIBioTic Granada 20F12/16. G.O thanks CONACyT Project No.~169125 for financial support. The authors acknowledge the Centro de Superc\'omputo UAM-Iztapalapa for the computational resources. 
\end{acknowledgments}


%

\end{document}